\documentclass[aps,prd,preprint,superscriptaddress,tightenlines,nofootinbib]{revtex4} 
\usepackage{epsfig}
\usepackage{graphicx}
\usepackage{dcolumn}
\usepackage{bm}

\def\upsl1{ \Upsilon {\rm (1S)}}

\def\ups4s{ \Upsilon (4S)}
\def\psidp{ \psi(3770) }
\def\ipb{{\rm pb^{-1}}}
\def\mbc{M_{\rm bc}}

\def\de{\Delta E}

\begin{document}
\preprint{\tighten\vbox{\hbox{\hfil CLNS 05-1943}
                        \hbox{\hfil CLEO 05-29}
}}


\title{
New Measurements of Cabibbo-Suppressed Decays of $D$ Mesons in CLEO-c}

\author{P.~Rubin}
\affiliation{George Mason University, Fairfax, Virginia 22030}
\author{C.~Cawlfield}
\author{B.~I.~Eisenstein}
\author{I.~Karliner}
\author{D.~Kim}
\author{N.~Lowrey}
\author{P.~Naik}
\author{C.~Sedlack}
\author{M.~Selen}
\author{E.~J.~White}
\author{J.~Wiss}
\affiliation{University of Illinois, Urbana-Champaign, Illinois 61801}
% ...
\collaboration{CLEO Collaboration} 
\noaffiliation

\date{Dec. 22, 2005}

\begin{abstract}
	Using 281 $\ipb$ of data collected with the CLEO-c detector,
we report on first observations and new measurements of Cabibbo-suppressed 
decays of $D$~mesons to 2, 3, 4, and 5 pions. Branching fractions
of previously unobserved modes are measured to be:
${\cal{B}}(D^0\to\pi^+\pi^-\pi^0\pi^0)=(9.9\pm0.6\pm0.7\pm0.2\pm0.1)\times10^{-3}$, 
${\cal{B}}(D^0\to\pi^+\pi^+\pi^-\pi^-\pi^0)=(4.1\pm0.5\pm0.2\pm0.1\pm0.0)\times10^{-3}$, 
${\cal{B}}(D^+\to\pi^+\pi^0\pi^0)=(4.8\pm0.3\pm0.3\pm0.2)\times10^{-3}$, 
${\cal{B}}(D^+\to\pi^+\pi^+\pi^-\pi^0)=(11.6\pm0.4\pm0.6\pm0.4)\times10^{-3}$,
${\cal{B}}(D^0\to\eta\pi^0)=(0.62\pm0.14\pm0.05\pm0.01\pm0.01)\times10^{-3}$, and
${\cal{B}}(D^0\to\omega\pi^+\pi^-)=(1.7\pm0.5\pm0.2\pm0.0\pm0.0)\times10^{-3}$.
The uncertainties are from statistics, experimental systematics, normalization and 
$CP$ correlations (for $D^0$ modes only). Improvements in other multi-pion
decay modes are also presented.
The $D\to\pi\pi$ rates allow us to extract the ratio of isospin amplitudes 
$A(\Delta I=3/2)/A(\Delta I=1/2)=0.420\pm0.014({\rm stat})\pm0.016({\rm syst})$ 
and the strong phase shift of $\delta_I=(86.4\pm2.8\pm3.3)^{\circ}$, which is
quite large and now more precisely determined. 
\end{abstract}

\maketitle

\setcounter{footnote}{0}

\newpage

	Charm decays provide an important laboratory for the
study of both the strong and weak interactions. Semileptonic decays provide
direct access to CKM matrix elements. Hadronic decays provide important input
to $B$ physics as well opening a window into
the study of final state (strong) interactions. Exhaustive or precise measurements
of Cabibbo-suppressed decays have been challenging due to low rates or other 
experimental challenges, however, more information on these decays is of great importance in
several areas. Firstly, in the weak sector, Cabibbo-suppressed final states, 
such as $D^0\to\pi^+\pi^-\pi^0$, provide a promising way by which to
extract the CKM angle $\gamma$ in $B^+\to D^{(*)0}K^{(*)}$ due to the
similar magnitude of interfering amplitudes between $D^0\to\pi^+\pi^-\pi^0$
and $\bar{D^0}\to\pi^+\pi^-\pi^0$~\cite{gamma}. In the arena of strong interactions,
they provide a means in which to study final state 
interactions~\cite{dpipi2,dpipi3}, which may help clarify how
they contribute to $B\to\pi\pi$ decay~\cite{dpipi1}.
Moreover these modes can enter as backgrounds to other $D$ decay 
measurements~\cite{dback} and therefore their values are 
of general importance to ascertain. Information on these decays is sparse
and not very precise. The CLEO-c $\psidp\to D\bar{D}$ sample provides
the opportunity to perform a comprehensive and precise study of these
decays.

	The CLEO-c detector is a general purpose solenoidal detector
which includes a tracking system for measuring momenta and specific ionization
of charged particles, a Ring Imaging Cherenkov detector to aid in particle
identification, and a CsI calorimeter for detection of electromagnetic showers.
The CLEO-c detector is described in detail elsewhere~\cite{cleo3}.
	
	This analysis utilizes 281 $\ipb$ of data collected on the $\psidp$
resonance ($\sqrt{s}\approx 3.77$ GeV) at the Cornell Electron Storage Ring. At
this energy, $D\bar{D}$ pairs are produced in a coherent $1^{--}$ final state
with no additional particles. 

	We reconstruct $D^0$~mesons in several multi-pion decay channels, including
$D^0\to\pi^+\pi^-$, $D^0\to\pi^0\pi^0$, 
$D^0\to\pi^+\pi^-\pi^0$, $D^0\to\pi^+\pi^+\pi^-\pi^-$, 
$D^0\to\pi^+\pi^-\pi^0\pi^0$,
$D^0\to\pi^+\pi^+\pi^-\pi^-\pi^0$ and $D^0\to\pi^0\pi^0\pi^0$. In 
$D^+$ we reconstruct $D^+\to\pi^+\pi^0$, $D^+\to\pi^+\pi^+\pi^-$,
$D^+\to\pi^+\pi^0\pi^0$, $D^+\to\pi^+\pi^+\pi^-\pi^0$, 
$D^+\to\pi^+\pi^+\pi^+\pi^-\pi^-$. We also consider the resonant $\eta$
and $\omega$ contributions and measure rates for
$D^0\to\eta\pi^0$, $D^0\to\omega\pi^+\pi^-$, and $D^+\to\eta\pi^+$.
The branching fractions are measured relative to $D^0\to K^-\pi^+$ and $D^+\to K^-\pi^+\pi^+$
for neutral and charged $D$~mesons, respectively. Throughout the paper,
charge conjugate modes are implicitly assumed, unless otherwise noted.

	The reconstruction of $D$~mesons uses charged particles 
and $\pi^0$'s reconstructed with standard selection requirements
~\cite{dhad}. For each candidate, we utilize
two kinematic variables: $\de=E_{\rm beam}-E_{D}$ and
$\mbc=\sqrt{E_{\rm beam}^2-p_{D}^2}$, where $E_{D}$ 
($p_{D}$) are the energy (momentum) of the $D$ candidate
and $E_{\rm beam}$ is the beam energy. The substitution of the beam energy for
the candidate energy improves the mass resolution by about
a factor of five. For properly reconstructed candidates,
$\de$ exhibits a narrow peak near zero and $\mbc$ peaks at the $D$
meson mass. Mode-dependent signal regions in $\de$ are defined to 
be within about three times the r.m.s. widths of the $\de$ distribution 
obtained from detailed Monte Carlo (MC) simulations 
(see Table~\ref{tab:tab1}). A $\de$ sideband region 
extending from 10~MeV beyond the signal region to 100~MeV is used to 
study the shape of the background in $\mbc$.

	The Cabibbo-favored modes such as $D\to K^0_S+X$, $K^0_S\to\pi\pi$
constitute a large background as their branching fractions are
typically five to ten times larger than the Cabibbo-suppressed modes, and 
therefore such events must be vetoed. For each decay channel with $\ge$ three pions,
we veto any candidate which contains a pair of pions with invariant mass in the range
$475<M_{\pi^+\pi^-}<520$~MeV/$c^2$ or $448<M_{\pi^0\pi^0}<548$~MeV/$c^2$. This
range was determined from a large sample of generic MC events where
we require that the surviving $K^0_S$ background be less than 1\% of the
expected signal in the corresponding Cabibbo-suppressed signal channel.

\begin{figure}
\centerline{
\includegraphics[width=3.5in]{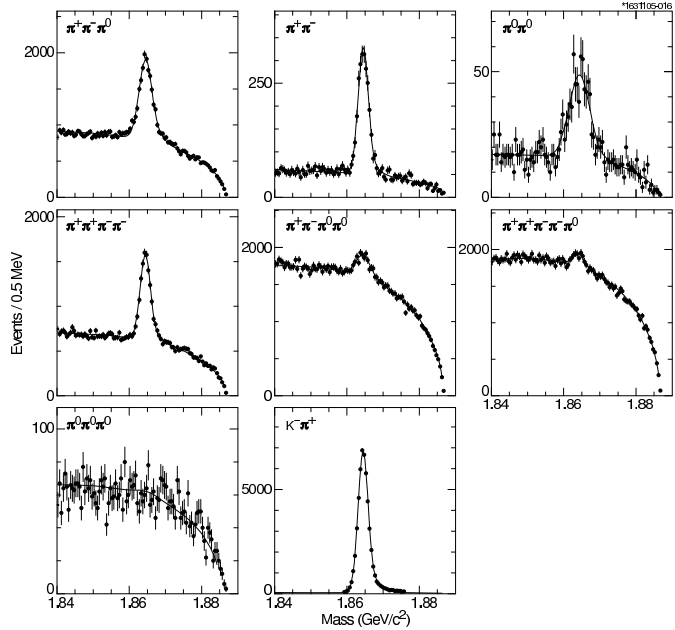}}
\caption{\label{fig:d0_data} 
$\mbc$ distributions for $D^0$ modes from data. The points are the
data and the superimposed lines are the fits as described in the text.}
\end{figure}

\begin{figure}
\centerline{
\includegraphics[width=3.5in]{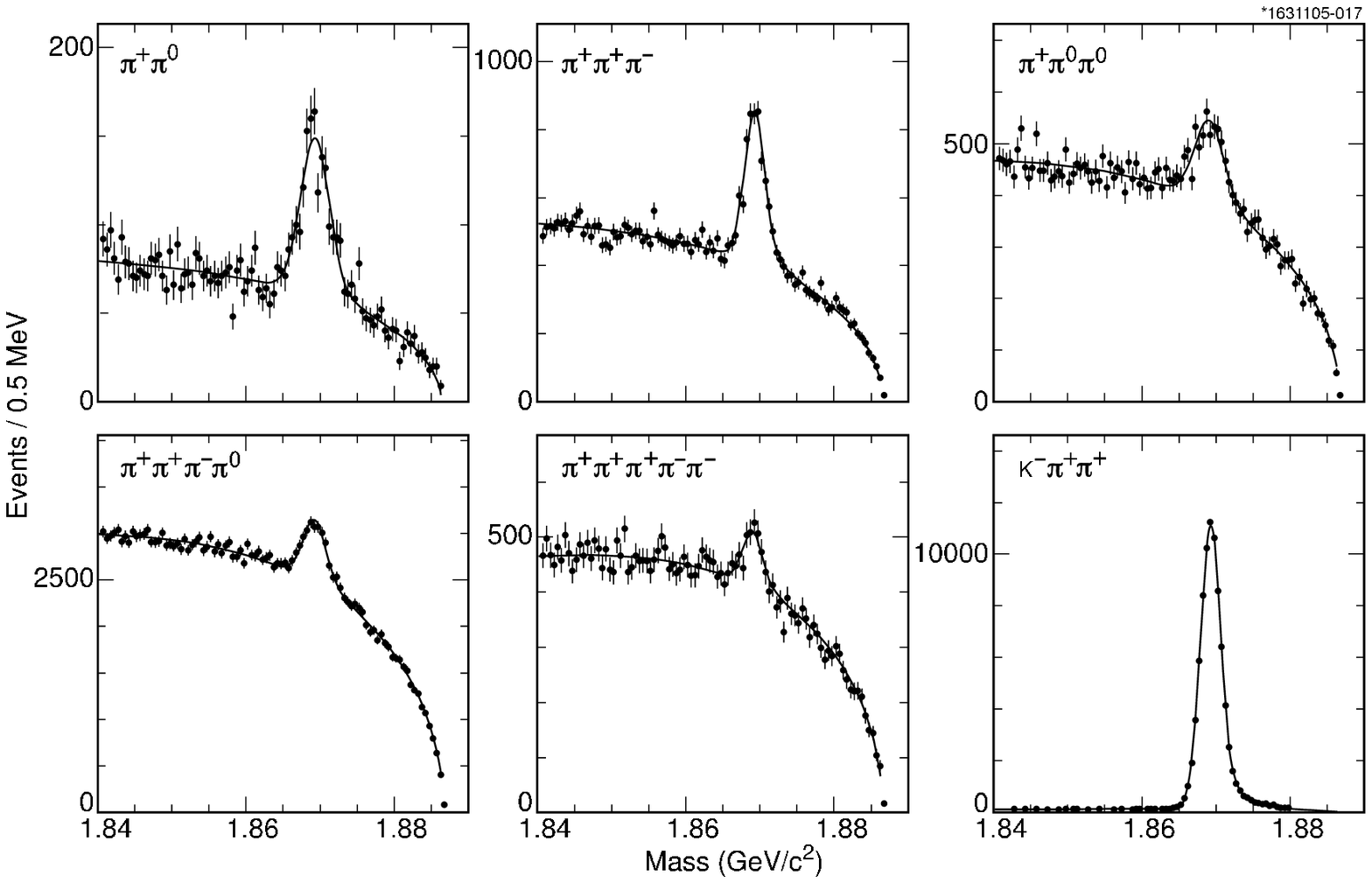}}
\caption{\label{fig:dch_data} 
$\mbc$ distributions for $D^+$ modes from data. The points are the
data and the superimposed lines are the fits as described in the text.}
\end{figure}

	The resulting $\mbc$ distributions for the neutral and charged
$D$ modes are shown in Figs.~\ref{fig:d0_data} and \ref{fig:dch_data},
respectively. The points show the data from the $\de$ signal region and the
lines are fits to the distributions which are given by the sum of an 
ARGUS threshold function~\cite{argus} and an
asymmetric signal shape(CBAL)~\cite{cbal}  which models the initial state 
radiation (ISR) induced tail at large $\mbc$. The {\sc argus} shape parameters 
are extracted by fitting the $\de$ sidebands. The signal
shape parameters are determined by fits to $\mbc$ distributions
obtained from a simulation of $D\bar{D}$ production
at the $\psi(3770)$~\cite{evtgen}
followed by a simulation of the detector response~\cite{geant,photos}. 
The Gaussian widths of the $\mbc$ distributions, which are part of the 
CBAL signal shape, and the fitted yields are listed in Table~\ref{tab:tab1}. 
The yields we find for the
normalization modes are in good agreement with the previously published
values~\cite{dhad}.

\begin{table}[hbt] 
\begin{center}  
\caption{Requirements on $\de$ for signal candidates,
Gaussian widths of the $\mbc$ distributions, observed yields, and 
reconstruction efficiencies. See text for details. \label{tab:tab1}}
\begin{tabular}{lcccc}\hline\hline 
Mode                            & $\de$    & $\mbc$     &               & Efficiency \\
	           		& [low, high] & Width      &  Yield        &    (\%)    \\    
				&  (MeV)   &   (MeV)      &  		&             \\
\hline
$D^0\to\pi^+\pi^-$ 		&  [-30, 30]     &  1.42      &  2085$\pm$54     &  73.1$\pm$0.7 \\
$D^0\to\pi^0\pi^0$ 		&  [-50, 40]     &  2.88      &  499$\pm$32      &  31.0$\pm$0.7 \\
$D^0\to\pi^+\pi^-\pi^0$    	&  [-50, 40]     &  1.83      &  10834$\pm$164   &  39.9$\pm$0.7 \\
$D^0\to\pi^+\pi^+\pi^-\pi^-$	&  [-25, 25]     &  1.46      &  7331$\pm$130    &  48.4$\pm$0.6 \\	
$D^0\to\pi^+\pi^-\pi^0\pi^0$	&  [-40, 30]     &  2.02      &  2724$\pm$166    &  15.9$\pm$0.5 \\
$D^0\to\pi^+\pi^+\pi^-\pi^-\pi^0$& [-30, 20]     &  1.67      &  1614$\pm$171    &  18.4$\pm$0.6 \\
$D^0\to\pi^0\pi^0\pi^0$ 	&  [-60, 35]     &  3.00      &  29$\pm$15       & 13.4$\pm$0.7  \\
$D^0\to~K^-\pi^+$ 		&  [-29, 29]     &  1.42      &  51210$\pm$231   & 65.0$\pm$0.6  \\  
\hline
$D^+\to\pi^+\pi^0$ 		&   [-40, 35]     &  1.98     &  914$\pm$46      &  46.5$\pm$0.8  \\
$D^+\to\pi^+\pi^+\pi^-$ 	&   [-25, 25]     &  1.40     &  3303$\pm$95     &  62.0$\pm$0.8  \\
$D^+\to\pi^+\pi^0\pi^0$ 	&   [-30, 30]     &  1.99     &  1535$\pm$89     &  22.0$\pm$0.6  \\
$D^+\to\pi^+\pi^+\pi^-\pi^0$ 	&   [-30, 30]     &  1.58     & 5701$\pm$205     &  30.6$\pm$0.7  \\
$D^+\to\pi^+\pi^+\pi^+\pi^-\pi^-$ & [-15, 15]     &  1.38     &  732$\pm$77      &  27.6$\pm$0.6  \\
$D^+\to~K^-\pi^+\pi^+$ 		&   [-22, 22]     &  1.37     &   80381$\pm$290  &  54.4$\pm$0.5  \\
\hline\hline 
\end{tabular} 
\end{center} 
\end{table} 

	Efficiencies for a given decay mode depend mildly on the presence of 
intermediate resonances, but the dependence is increased to as large as 10\%
(relative) when the $K^0_S$ veto is included. To minimize this bias, we
tune the MC simulation by either using existing Dalitz-plot analyses, as for the
$\pi\pi\pi$ final states~\cite{pipipi0_dalitz,pipipi_dalitz}, or
by adding intermediate resonances to each mode in order to reproduce the 
observed substructure in the $\pi\pi$ invariant mass distribution. In
most cases, this requires introducing significant $\rho$ contributions.
Efficiencies from simulation are checked in data by comparing the numbers
of particles of each species in fully reconstructed events with 
the corresponding yields when their populations are inferred from
energy and momentum conservation~\cite{dhad}. The resulting
corrections are: (-3.9$\pm$2.0)\% per $\pi^0$, (-0.7$\pm$0.7)\% per $K^{\pm}$
and (-0.3$\pm$0.3)\% per $\pi^{\pm}$~\cite{dhad}. For $\pi^0$'s 
with momentum near $M_D/2$, the correction is $(-4.6\pm3.5)\%$.
The resulting efficiencies are
shown in the last column of Table~\ref{tab:tab1}. 

	These multi-pion final states are fed by intermediate resonances,
such as $\rho$, $\eta$, $\omega$, $\phi$, $f_0$, $f_2$, etc. We examine
the multipion final states for $\eta$, $\omega\to\pi^+\pi^-\pi^0$ only. 
The $\phi\to\pi^+\pi^-\pi^0$ intermediate state is not treated here since it
can be measured significantly better using 
$\phi\to K^+K^-$, and the observed rates are too low to improve on 
existing measurements. 

\begin{figure}
\centerline{
\includegraphics[width=3.5in]{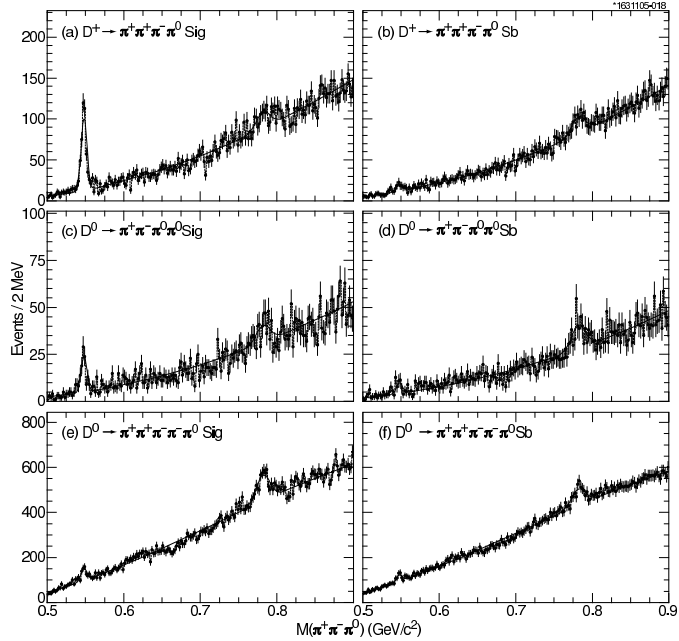}}
\caption{\label{fig:resmodes} 
Distributions in $\pi^+\pi^-\pi^0$ invariant mass for 
$D^+\to\pi^+\pi^+\pi^-\pi^0$ (a,b), $D^0\to\pi^+\pi^-\pi^0\pi^0$ (c,d),  
and $D^0\to\pi^+\pi^+\pi^-\pi^-\pi^0$ (e,f) candidates. The left column
represents the $\de$ signal region and the right are for $\de$
sidebands. The superimposed lines are the fits as described in the text.}
\end{figure}

	We search the $D^+\to\pi^+\pi^+\pi^-\pi^0$, 
$D^0\to\pi^+\pi^-\pi^0\pi^0$,  and $D^0\to\pi^+\pi^+\pi^-\pi^-\pi^0$ 
modes for $\eta$ and $\omega$ decays. The yields are extracted by
selecting events that are within 2.5 times the r.m.s. width of the $D$ mass
and taking the difference in yields between the number of such events 
in the $\de$ signal and $\de$ sideband regions.
The sideband distributions are normalized to account for the different
range of $\de$ between signal and sidebands regions. Distributions 
of $\pi^+\pi^-\pi^0$ mass for these
three modes are shown in Fig.~\ref{fig:resmodes}. The left column
shows combinations whose $\de$ is in the signal region and the right
column shows candidates in the $\de$ sideband region. The distributions
are fit to the sum of a polynomial background and two Gaussians, one
each for the $\eta$ and $\omega$. With the limited statistics and
large background, we do not try to fit the 
Breit-Wigner tails of the $\omega$, but rather assume a Gaussian shape and
absorb the loss of events from the tails into the efficiency. In
the fit, the $\eta$ and $\omega$ Gaussian widths are constrained to 3.8~MeV 
and 9.0~MeV, respectively, 
the values determined from signal MC. The resulting yields for the $\de$
signal and sideband regions are shown in Table~\ref{tab:tab2} along
with the reconstruction efficiency, determined from signal MC.

\begin{table}[hbt] 
\begin{center}  
\caption{Summary of submodes containing $\eta$ and $\omega$ mesons, 
indicating efficiencies and yields in the $\de$ signal and $\de$ sideband 
regions.\label{tab:tab2}}
\begin{tabular}{lcccccc}\hline\hline 
Mode 	 & Efficiency & Yield        & Yield      \\
    &          (\%)  	& $\de$ signal  & $\de$ sideband \\
\hline
$D^+\to\eta\pi^+$ & 29.4$\pm$1.0 & 421$\pm$23 & 44$\pm$12  \\
$D^+\to\omega\pi^+$ & 21.7$\pm$1.0 & 216$\pm$43 & 236$\pm$41 \\
$D^0\to\eta\pi^0$ & 21.9$\pm$1.0 & 90$\pm$12 & 28$\pm$8 \\
$D^0\to\omega\pi^0$ & 14.1$\pm$1.0 & 103$\pm$26 & 140$\pm$25 \\
$D^0\to\eta\pi^+\pi^-$&20.3$\pm$1.0 & 260$\pm$32 & 150$\pm$29 \\
$D^0\to\omega\pi^+\pi^-$&15.6$\pm$1.1 & 1304$\pm$96 & 832$\pm$91 \\
\hline\hline 
\end{tabular} 
\end{center} 
\end{table} 

	Relative branching fractions are computed using the yields 
shown in Tables~\ref{tab:tab1} and \ref{tab:tab2} and are presented in 
Table~\ref{tab:tab3} (unseen decay modes of the $\eta$ and $\omega$
are included, using ${\cal{B}}(\eta,\omega\to\pi^+\pi^-\pi^0)$ 
from Ref.~\cite{pdg}). To compute the absolute branching fractions, we
use ${\cal{B}}(D^0\to K^-\pi^+)=(3.84\pm0.07)\%$ and
${\cal{B}}(D^+\to K^-\pi^+\pi^+)=(9.4\pm0.3)\%$, which are obtained
using a weighted average of the PDG values and the recent CLEO 
measurements~\cite{dhad}. For unobserved modes, we set 90\% confidence level 
upper limits. In the last column of Table~\ref{tab:tab3} 
we show PDG~\cite{pdg} averages, when available. 

\begin{table*}[hbt] 
\begin{center}  
\caption{Measured relative and absolute branching fractions for neutral and charged 
$D$ modes. Uncertainties are statistical, experimental systematic, 
normalization mode uncertainty, and uncertainty from $CP$ correlations 
(for $D^0$ modes only). For the relative branching fractions, the
normalization mode uncertainty is omitted.
\label{tab:tab3}}
\begin{tabular}{lccc}\hline\hline 
Mode 		& ${\cal{B}}_{\rm mode}/{\cal{B}}_{\rm ref}$ & ${\cal{B}}_{\rm mode}$  & ${\cal{B}}$ (PDG) \\
                &                (\%)                       &   ($10^{-3}$)   & ($10^{-3}$) \\
\hline
$D^0\to\pi^+\pi^-$ & $~~~3.62\pm0.10\pm0.07\pm0.04~~~$ & $~~~1.39\pm0.04\pm0.04\pm0.03\pm0.01~~~~$ & 1.38$\pm$0.05 \\
$D^0\to\pi^0\pi^0$ & $2.05\pm0.13\pm0.16\pm0.02$  & $0.79\pm0.05\pm0.06\pm0.01\pm0.01$ & 0.84$\pm$0.22 \\
$D^0\to\pi^+\pi^-\pi^0$ & $34.4\pm0.5\pm1.2\pm0.3$ & $13.2\pm0.2\pm0.5\pm0.2\pm0.1$ & 11$\pm$4 \\
$D^0\to\pi^+\pi^+\pi^-\pi^-$ & $19.1\pm0.4\pm0.6\pm0.2$ & $7.3\pm0.1\pm0.3\pm0.1\pm0.1$ & 7.3$\pm$0.5 \\
$D^0\to\pi^+\pi^-\pi^0\pi^0$ & $25.8\pm1.5\pm1.8\pm0.3$ & $9.9\pm0.6\pm0.7\pm0.2\pm0.1$ &  \\ 
$D^0\to\pi^+\pi^+\pi^-\pi^-\pi^0$ & $10.7\pm1.2\pm0.5\pm0.1$ & $4.1\pm0.5\pm0.2\pm0.1\pm0.0$ &  \\
$D^0\to\omega\pi^+\pi^-$& $4.1\pm1.2\pm0.4\pm0.0$ &  $1.7\pm0.5\pm0.2\pm0.0\pm0.0$ &  \\
$D^0\to\eta\pi^0$  &  $1.47\pm0.34\pm0.11\pm0.01$ & $0.62\pm0.14\pm0.05\pm0.01\pm0.01$ &  \\
$D^0\to\pi^0\pi^0\pi^0$ &  -  & $<0.35$ (90\% CL) &  \\
$D^0\to\omega\pi^0$ &  -  & $<0.26$ (90\% CL) &   \\
$D^0\to\eta\pi^+\pi^-$ &  -  & $<1.9$ (90\% CL) &   \\
\hline
$D^+\to\pi^+\pi^0$ & 1.33$\pm$0.07$\pm$0.06 & 1.25$\pm$0.06$\pm$0.07$\pm$0.04 & 1.33$\pm$0.22 \\
$D^+\to\pi^+\pi^+\pi^-$ & 3.52$\pm$0.11$\pm$0.12 & 3.35$\pm$0.10$\pm$0.16$\pm$0.12 & 3.1$\pm$0.4 \\
$D^+\to\pi^+\pi^0\pi^0$ & 5.0$\pm$0.3$\pm$0.3 & 4.8$\pm$0.3$\pm$0.3$\pm$0.2 &  \\
$D^+\to\pi^+\pi^+\pi^-\pi^0$ & 12.4$\pm$0.5$\pm$0.6 & 11.6$\pm$0.4$\pm$0.6$\pm$0.4&  \\
$D^+\to\pi^+\pi^+\pi^+\pi^-\pi^-$ & 1.73$\pm$0.20$\pm$0.17 & 1.60$\pm$0.18$\pm$0.16$\pm$0.06 & 1.73$\pm$0.23 \\
$D^+\to\eta\pi^+$ & 3.81$\pm$0.26$\pm$0.21 & 3.61$\pm$0.25$\pm$0.23$\pm$0.12 & 3.0$\pm$0.6 \\
$D^+\to\omega\pi^+$ &  -  & $<0.34$ (90\% CL) &   \\
\hline\hline
\end{tabular} 
\end{center} 
\end{table*}

	Several sources of systematic uncertainty are considered. Where 
appropriate, we include in parentheses the minimum to maximum size of the uncertainty.
Limited MC statistics
are used in determining the reconstruction efficiencies, which introduce relative
uncertainties at the level of (1-3)\%. Tracking and $\pi^0$ reconstruction 
efficiencies have relative uncertainties of 0.7\% and 2.0\% (3.5\% for high momentum $\pi^0$)
per particle, 
respectively, and charged particle identification is understood at the level 
0.3\%/$\pi^{\pm}$ and 1.3\%/$K^{\pm}$. Uncertainty from the $K^0_S$ veto is 
estimated as the difference in probabilities (between data and MC) for each final
state to pass the $K^0_S$ veto. This survival fraction, $F^{\pi\pi}$, is given by
$F^{\pi\pi} \equiv (1-f_{\rm veto})^{N_{\rm pair}}$, where $f_{\rm veto}$ represents the veto 
probability per $\pi\pi$ pair, which is obtained by a linear interpolation from
an 80~MeV region above and below the veto region, and $N_{\rm pair}$ is the number
of $\pi^+\pi^-$ (or $\pi^0\pi^0$) pairs in the given decay mode. The uncertainties on the
veto efficiencies for $\pi^+\pi^-$ and $\pi^0\pi^0$ are added in quadrature
when applicable (0.0\%-2.8\%). Uncertainties in signal yields are estimated by comparing
changes in the branching fraction when (a) signal widths are permitted to float,
(b) varying the ISR-tail shape parameters individually by $\pm 1$ standard deviation, and 
(c) varying the range of the fit to the $\mbc$ distribution. The three sources 
are added in quadrature to obtain the total uncertainties (1.0\%-4.9\%). Uncertainties
resulting from different $\de$ resolutions are quantified by widening the
$\de$ windows, recomputing the branching fractions, and taking the difference
between the new and default values (0.2\%-7.7\%). The simulation of final state radiation
has been studied in $J/\psi\to\mu^+\mu^-$ events and is estimated to introduce no more
than 0.5\% uncertainty on the $D$ reconstruction efficiency. Possible bias
due to resonant substructure is estimated by removing the $K^0_S$ veto (so as not
to double count this systematic) and comparing the efficiencies obtained
using our default simulation (which includes resonances) and a phase space 
simulation (0.0\%-2.2\%).
Imperfect replication of the average number of candidates per event by the simulation
could lead to a bias in the reconstruction efficiency. This effect is quantified by comparing the
average number of $D$ candidates per event, $\langle N_{\rm cand}\rangle$, within 2.5 standard deviations of the $D$ mass,
between data and simulation (on a mode-by-mode basis). The systematic uncertainty
is taken as the difference in the ratio from unity, {\it i.e.,} 
$\langle N_{\rm cand}^{\rm data}\rangle /\langle N_{\rm cand}^{\rm MC}\rangle -1$
(0.2\%-2.7\%). The $D^0$ and $D^+$ normalization modes introduce uncertainties of 1.8\% and 
3.2\%, respectively. Lastly, the effects of quantum correlations of $D\bar{D}$
pairs produced from the $1^{--}$ $\psi(3770)$ may shift
the branching fractions with respect to the values in the absence of
these correlations. The effect of these quantum correlations has been
considered~\cite{asner_sun_cp}, and a shift in the branching fraction
can be expected if $y=\Delta\Gamma/\Gamma\neq~0$. Limits on $y$ are at the
percent level~\cite{pdg}, and we take this as an additional systematic 
uncertainty on the $D^0$ branching fractions.
The total systematic uncertainties on the branching fractions for 
each mode are shown in Table~\ref{tab:tab3}.


	Using the new $D\to\pi\pi$ branching fractions and $D$ lifetimes,
$\tau_{D^+}=(1040\pm7)$ fs and $\tau_{D^0}=(410.3\pm1.5)$ fs~\cite{pdg}, we compute 
the ratio of the $\Delta I=3/2$ to $\Delta I=1/2$ isospin amplitudes 
and their relative strong phase difference~\cite{isospin} to be 
$A_2/A_0=0.420\pm0.014({\rm stat})\pm0.01({\rm syst})$ and
$\cos\delta_I=0.062\pm0.048({\rm stat})\pm0.058({\rm syst})$. 
The large phase shift, $\delta_I=(86.4\pm2.8\pm3.3)^{\circ}$, indicates
that final state interactions are important in $D\to\pi\pi$ transitions. 
These results represent a considerable improvement over
previous measurements from CLEO~\cite{cleo-pipi2} and FOCUS~\cite{focus},
both of which are consistent with our data.
	
	In summary, we report new measurements of six Cabibbo-suppressed
decay branching fractions of $D$ mesons and eight additional 
measurements, of which all except for $D^0\to\pi^+\pi^-$ and
$D^+\to\pi^+\pi^+\pi^+\pi^-\pi^-$ 
provide large improvements over
the existing world average values. The large value of the strong phase
shift, $\delta_I$ obtained in the $D\to\pi\pi$ decay 
supports the conclusion 
that final state interactions are important in $D$ decays.

We gratefully acknowledge the effort of the CESR staff 
in providing us with excellent luminosity and running conditions.
This work was supported by the A.P.~Sloan Foundation,
the National Science Foundation, and the U.S. Department of Energy.



\begin{thebibliography}{99}
\bibitem{dback}
For example, $D^+\to\pi^+\pi^0$ is a background to $D^+\to\mu^+\nu_{\nu}$,
$D^+\to\pi^+\pi^0\pi^0$ is a background to the inclusive $D^+\to K^0\pi^+$
analysis.
\end{thebibliography}
\end{document}